\documentclass[prd,preprintnumbers,nofootinbib,superscriptaddress,onecolumn,notitlepage]
{revtex4-1}

\pdfoutput=1
\usepackage{graphicx}
\usepackage{amsmath}
\usepackage{amssymb}
\usepackage{slashed}
\usepackage{braket}
\usepackage{mathtools}
\usepackage[utf8]{inputenc}
\usepackage[colorlinks=true,linkcolor=blue,bookmarksopen,bookmarksnumbered,citecolor=magenta]{hyperref}
\usepackage[usenames,dvipsnames]{xcolor}
\usepackage[normalem]{ulem}
\usepackage{soul}
\usepackage{units}
\usepackage{rotating}
\usepackage{hhline,multirow,tabularx}
\usepackage{bm}
\usepackage{longtable}
\usepackage{makecell}
\usepackage{setspace}

\newcommand{\T}[2]{\boldsymbol{#1}_{#2\text{T}}}
\newcommand{\THD}[1]{\boldsymbol{#1}_{H,\text{T}}}

\newcommand{\Tsc}[2]{#1_{#2\text{T}}}
\newcommand{\Tscsq}[2]{#1^2_{#2\text{T}}}

\newcommand{\parz}[1]{\ensuremath{\left(#1\right)}}
\newcommand{\order}[1]{\ensuremath{O\parz{#1}}}

\newcommand{\eref}[1]{Eq.~(\ref{e.#1})}
\newcommand{\erefs}[2]{Eqs.~(\ref{e.#1})--(\ref{e.#2})}
\newcommand{\fref}[1]{Fig.~\ref{f.#1}}

\newcommand{\sref}[1]{Sec.~\ref{s.#1}}

\newcommand\BIBvol[1]{\textbf{#1}}

\definecolor{dpmagenta}{rgb}{0.8, 0.0, 0.8}

\allowdisplaybreaks
\DeclareRobustCommand*\diff[2][]{%
   \mathop{
     \mathrm{d}^{#1}
     \mskip-0.2\thinmuskip
    #2}\nolimits
}

\begin{document}

\title{Challenges with Large Transverse Momentum in Semi-Inclusive Deeply Inelastic Scattering}

\preprint{JLAB-THY-18-2782}

\author{J.~O.~Gonzalez-Hernandez}
\email{joseosvaldo.gonzalez@to.infn.it}
\affiliation{Department of Physics, Old Dominion University, Norfolk, VA 23529, USA}
\affiliation{Dipartimento di Fisica, Universit\`a di Torino, Via P. Giuria 1, 10125 Torino, Italy}
\affiliation{INFN-Sezione Torino, Via P. Giuria 1, 10125 Torino, Italy}

\author{T.~C.~Rogers}
\email{trogers@jlab.org}
\affiliation{Department of Physics, Old Dominion University, Norfolk, VA 23529, USA}
\affiliation{Jefferson Lab, 12000 Jefferson Avenue, Newport News, VA 23606, USA}
\author{N.~Sato}
\email{nsato@jlab.org}
\affiliation{Jefferson Lab, 12000 Jefferson Avenue, Newport News, VA 23606, USA}
\author{B.~Wang}
\email{0617626@zju.edu.cn}
\affiliation{Department of Physics, Old Dominion University, Norfolk, VA 23529, USA}
\affiliation{Jefferson Lab, 12000 Jefferson Avenue, Newport News, VA 23606, USA}
\affiliation{Zhejiang Institute of Modern Physics, Department of Physics, Zhejiang University,\\ Hangzhou 310027, China}

\date{12 December 2018}

\begin{abstract}
We survey the current phenomenological status of semi-inclusive deeply inelastic 
scattering at moderate hard scales and in the limit of very large transverse momentum. As 
the transverse momentum becomes comparable to or larger than the overall hard scale, 
the differential cross sections should be calculable with fixed order perturbative (pQCD) methods, while 
small transverse momentum (TMD factorization) approximations should eventually break 
down. We find large disagreement between HERMES and COMPASS data and fixed 
order calculations done with modern parton densities, even in regions of kinematics 
where such calculations should be expected to be very accurate. Possible interpretations 
are suggested.
\end{abstract}

\maketitle

\section{Introduction}
\label{s.intro}
Transverse momentum spectra are of theoretical interest for many reasons, and processes 
with an electromagnetic hard scale like Drell-Yan scattering (DY) and semi-inclusive 
deeply inelastic scattering (SIDIS) are ideal clean probes of the underlying hadronic 
correlation functions. In many efforts dedicated specifically to probing the details 
of hadronic structure, the hard scales involved are relatively low or moderate, making a 
distinction between the different regions delicate. 

In this article, we will focus on SIDIS, $$l(l) + \text{Proton}(P) \to l^{\prime}(l^{\prime})+
\text{Hadron}(P_H) + X,$$ wherein a single identified hadron with momentum $P_H$ is 
observed in the final state. The virtuality of the space-like momentum $q\equiv l^{\prime}-l$ is used to define 
a hard scale $Q\equiv \sqrt{-q^2}$ for the process. For us, the phrase ``transverse momentum'' is $\T{q}{}$, 
the transverse momentum of the virtual photon in a frame where $P$ and $P_H$ are back-
to-back. (See \sref{partonkin} for a detailed overview of our notation.)
When the SIDIS cross section is differential in both $\Tsc{q}{}$ and $z$, it displays the 
relative contributions from different underlying physical 
mechanisms for transverse momentum generation.  For the purposes of this article, we are 
interested in cases where $z$ is large enough 
to be in the current fragmentation region, wherein it can be associated with a fragmentation 
function (FF). Then there are a further three transverse momentum subregions, each with 
its own physical interpretation:

\begin{enumerate}
	\item When the transverse momentum is very small (between $0$ and a scale of 	
	order a 
	hadron mass), it is usually understood to have been generated by non-perturbative 
	processes intrinsic to the incoming proton or outgoing measured hadron. 
	This is the transverse momentum dependent (TMD) factorization region, and it has 
	attracted major attention in recent years due to its connection to intrinsic non-	
	perturbative properties of partons inside hadrons.  
	(See Refs.~\cite{Collins:1981uw,Collins:1981uk,
	Collins:1981tt,Collins:1984kg} for the Collins-Soper-Sterman (CSS) formalism, and
	Refs~\cite{Collins:2011qcdbook} for the updated version of TMD
	factorization by Collins. See a recent review, Ref.~\cite{Rogers:2015sqa}, by one of 
	us, which contains more references. For approaches rooted in soft-collinear effective 
	theory, see especially Refs.~\cite{GarciaEchevarria:2011rb,Echevarria:2016scs,
	Luebbert:2016itl,Li:2016axz,Li:2016ctv,Scimemi:2017etj,Gutierrez-Reyes:2018qez}.)
	TMD factorization theorems apply to the limit of $\Tsc{q}{} \ll Q$ since neglected 
	terms are suppressed by powers of $\Tscsq{q}{}/Q^2$.
	\item At still small but somewhat larger transverse momenta, there is a regime where 
	$\Lambda_{\rm QCD}^2 \ll \Tscsq{q}{} \ll Q^2$. The ratio $\Tscsq{q}{}/Q^2$ continues 
	to be small, so the same TMD factorization methods of region 1 continue to apply. 	
	However, $\Lambda_{\rm QCD}^2/\Tscsq{q}{}$ is also small, so the transverse 
	momentum in this region may be described largely by perturbative or 
	semi-perturbative techniques, often in the form of resummed logarithms of transverse 
	momentum. The general concepts of TMD parton distribution functions (PDFs) and 
	TMD FFs remains valid, and the transition between regions 1 and 2 happens 
	naturally as part of a general TMD factorization formalism. Therefore, it is reasonable 
	in many contexts to just treat them as a single region, as is usually done.
 	\item However, at even larger transverse momenta where $\Tsc{q}{} \gtrsim Q$ the 
	 $\Tscsq{q}{}/Q^2$-suppressed terms that are neglected in TMD factorization are not 
	 necessarily negligible. (Indeed, the logarithms induced by the small $\Tsc{q}{}/Q$ 	
	 approximations can create large errors in the very large $\Tsc{q}{}$ regions.) In this 	
	 region, the transverse momentum is probably best understood not as an intrinsic 
	 property of the hadrons, but instead as something produced directly in a 
	 process-dependent hard collision. Fortunately, in this situation there are two valid and 
	 comparable hard scales, $Q$ and $\Tsc{q}{}$, so fixed order calculations with 
	 pure collinear factorization should be very reliable.  This most direct pQCD approach 
	 begins to fail if $\Tsc{q}{}$ is too small to play the role of a hard scale comparable to $Q$, and this shows 
	 up in fixed order calculations as terms that diverge as 
	 $\Tsc{q}{}/{Q} \to 0$. In that limit, one must return to the methods of regions 1 or 2.
 \end{enumerate}
Since regions 2 and 3 both deal with the limit of $\Tsc{q}{} \gg \Lambda_{\rm QCD}$, they 
both might reasonably be referred to as ``large transverse momentum'' regions. However, 
it is important to keep the distinction between them clear, particularly for this paper. The 
former uses the $\Tsc{q}{}/Q \ll 1$ approximations of TMD factorization while the latter 
does not. In this paper, ``large transverse momentum'' will always refer specifically to 
region 3. See~\cite{Bacchetta:2008xw} for more details on the matching between different 
types of behavior at large and small transverse momentum and~\cite{Anselmino:2006rv} for an 
early phenomenological perspective.

SIDIS is fully understood only after each of these three subregions is understood on its 
own and only after it is clear how they match onto one another for the full range of 
kinematical scales from small to large $Q$. Most especially, identifying properties of 
transverse momentum dependence that are truly intrinsic to specific hadrons requires that
they be disentangled from those that are generated in process-specific hard collisions. 
This can be delicate, especially at the smaller values of $Q$ typical of SIDIS experiments, 
because the three regions enumerated above begin to be squeezed into an increasingly 
small range of $\Tsc{q}{}$ (see \cite{Boglione:2014oea,Collins:2017ybb}).  

Of course, the above classification of transverse momentum regions is not specific to 
SIDIS. In fact, the more common introduction to the subject of transverse momentum 
dependence in pQCD and its physical origins usually begins by considering processes like 
DY scattering. In the standard introduction, region 3 styles of calculation appear to be the 
more manageable scenario, given that all scales are both 
large and comparable so that asymptotic freedom applies and there are none of the 
diverging logarithms associated with $\Tsc{q}{} \to 0$ limit. (See, for example, chapter 5.6 
of \cite{Field:1989uq} and chapter 9.1 of \cite{Ellis:1996qj}.) As long as both the $l^+ l^-$ 
mass $Q$ and the center-of-mass transverse momentum $\Tsc{q}{}$ are comparably 
large, one expects these calculations to be at least very roughly consistent with 
measurements. 

Away from very large hard scales (such as weak boson mass scales, $Q \gtrsim 80$~GeV) 
the number of more recent phenomenological studies designed specifically to test region 3 
calculations on their own merits is surprisingly small. But understanding the transition 
to the $\Tsc{q}{} \sim Q$ region is important for clarifying the general nature of transverse 
momentum dependence in processes like SIDIS, especially for more moderate values of 
$Q$ where the transition between regions is not obvious. Furthermore, these are highly 
constrained calculations since the only input objects that involve prior fitting -- the PDFs 
and FFs -- are those taken from collinear factorization. Thus, they yield highly 
unambiguous predictions with no fitting parameters. 

However, as we will show, region 3 calculations that use modern PDF and FF sets do not 
in general produce even roughly successful predictions in SIDIS, even for values of $x$, 
$z$, $Q^2$ and transverse momentum where the expectation is that fixed order 
calculations should be very reliable. 

We will discuss further the delineation between different regions in \sref{partonkin}, where 
we will also explain our notation. In \sref{results}, we will show examples of large 
transverse momentum behavior compared with existing data and find 
that for moderate $x$, moderate $z$, $Q$ of a few GeVs, and $\Tsc{q}{} \gtrsim Q$, 
existing data are poorly described by both leading order or next-to-leading order 
calculations. In \sref{conc}, we comment on our observations.

\section{Factorization and regions of partonic kinematics}
\label{s.partonkin}

We will express quantities in terms of the conventional kinematical variables
$z \equiv P_H \cdot P/ (P \cdot q)$ and $x\equiv Q^2/2P\cdot q$. $\T{P}{H,}$ is the Breit 
frame transverse momentum of the produced hadron, and $P$ and $q$ are the 
four-momenta of the incoming target hadron and the virtual photon respectively.  
We assume that $x$ and $1/Q$ are small enough that both the proton, 
final state hadron, and lepton masses can be dropped in phase space factors. 
(A word of caution is warranted here, since the values of $Q$ for the experiments we examine here can be quite low. 
In the future, target and hadron mass effects should be examined in greater detail using methods such as those discussed in Refs.~\cite{Brady:2011uy,Guerrero:2015wha}.)
As mentioned in the introduction, it is useful to express transverse momentum in terms of 
\begin{equation}
\T{q}{} = -\frac{\THD{P}}{z} \, .  \label{e.qtdef}
\end{equation}
In a frame where the incoming and outgoing hadrons are back-to-back,
$\T{q}{}$ is the transverse momentum of the virtual photon. 
\begin{figure}
\centering
\centering
  \begin{tabular}{c@{\hspace*{5mm}}c}
    \multicolumn{2}{c}{
    \includegraphics[scale=0.45]{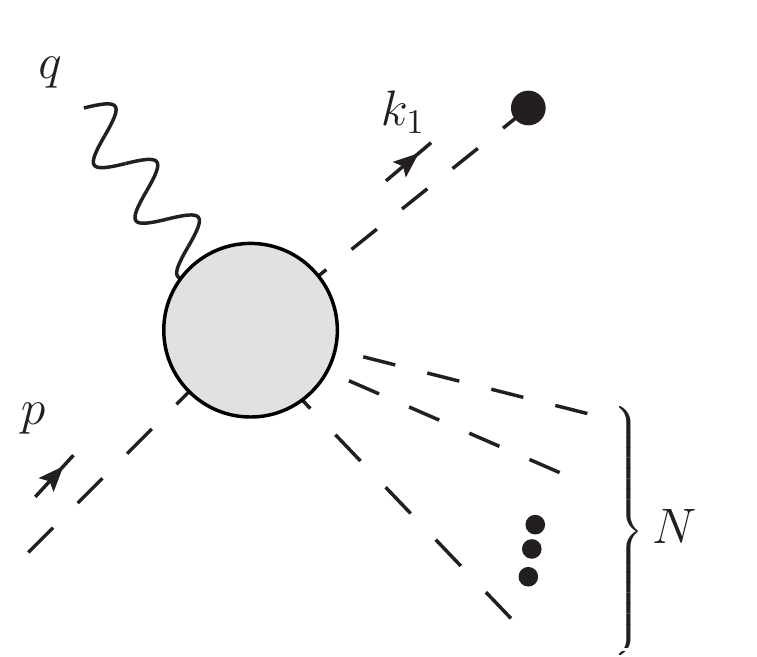}
  }
  \\
  \multicolumn{2}{c}{(a)} \\
    \includegraphics[scale=0.45]{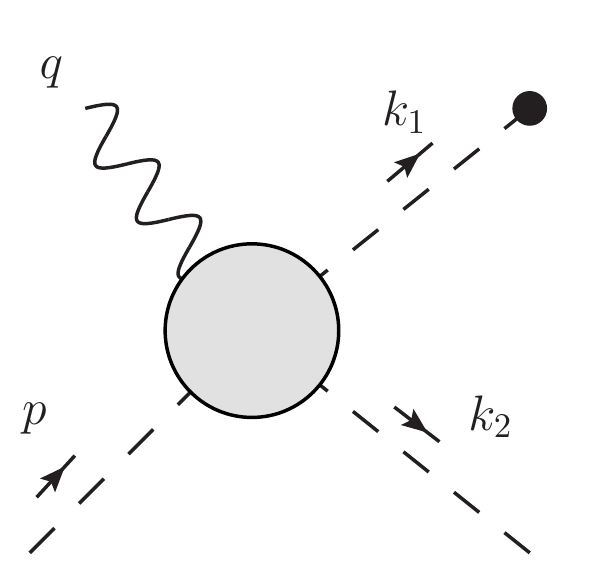}
    \hspace{0.5cm}
    &
    \hspace{0.5cm}
    \includegraphics[scale=0.45]{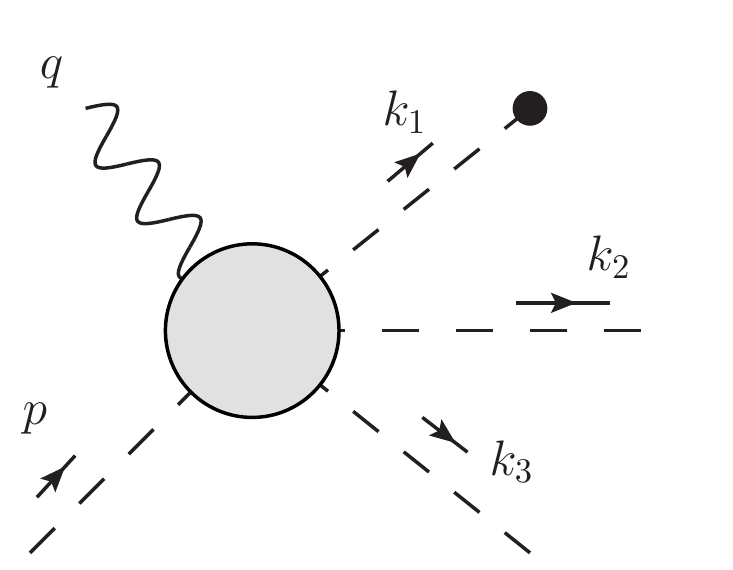}
  \\
  (b) & (c)
  \end{tabular}
\caption{Momentum labeling in amplitudes for $2 \to N$ partonic scattering kinematics. The 
dashed lines represent 
partons of unspecified type and flavor. The dot on the end of $k_1$ is to indicate 
this is the parent parton of the detected hadron. The other momenta are 
integrated in SIDIS. All final state lines are meant to represent energetic but mutually highly 
non-collinear massless partons. If two lines become nearly collinear, it should be 
understood that they merge into a single line. If a line becomes soft, it is simply to be 
removed. In both cases $2 \to N$ kinematics reduce to $2 \to N-1$ kinematics. 
(a) is a general $2 \to N$ amplitude. For perturbatively large $\Tsc{P}{H}$, $N \geq 1$ and 
the lowest order contribution is $\order{\alpha_s}$, corresponding to 
the $2\to2$ kinematics in (b). We will consider in addition the $2 \to 3$ kinematics 
in (c), which appear at $\order{\alpha_s^2}$. 
}
\label{f.basickinematics}
\end{figure}
\begin{figure}
\includegraphics[scale=0.5]{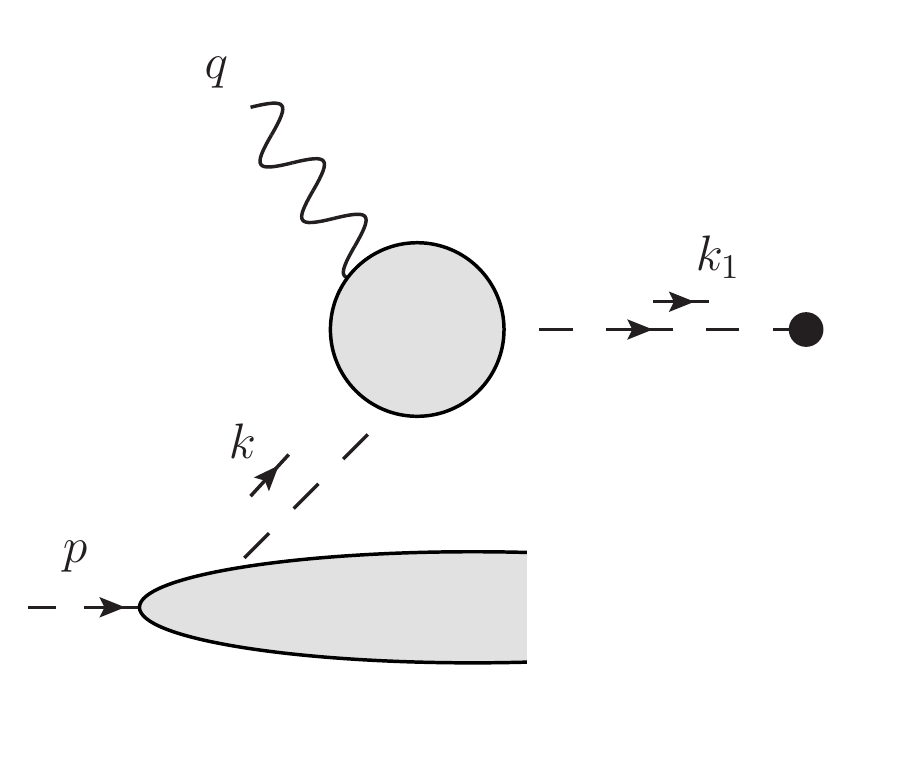}
\caption{Figure \ref{f.basickinematics}(a) reduces to this handbag structure when 
$|k^2| \sim \Lambda_{\rm QCD}^2$. The $k$ line becomes part of the target parton, and the 
lower blob is part of a PDF in the square-modulus amplitude integrated over final states.}
\label{f.hand}
\end{figure}

The factorization theorem that relates the hadronic and partonic differential cross sections 
in SIDIS at large $\THD{P}$ is
\begin{equation}
4 P_{\rm H}^0 E^\prime \frac{\diff{\sigma}{}}{\diff{^3 {\bf l}{'}} \, \diff{^3 {\bf P}_{H} }}  =
  \int_{x}^{1} \frac{\diff{\xi}}{\xi} 
  \int_{z}^{1} \frac{\diff{\zeta}}{\zeta^2} 
  \left( 4 k_1^0 E^\prime \frac{\diff{\hat{\sigma}_{ij}}{}}{\diff{^{3} {\bf l}{'}} \, \diff{^{3} {\bf k}
  _1 }} \right) f_{i/P}(\xi;\mu) d_{H/j}(\zeta;\mu) \, 
+ O(\Lambda_{\rm QCD}^2/q_{\rm T}^2) \, . \label{e.xsecfact}
\end{equation}
The $1/\xi$ is from the partonic flux factor, and the 1$/\zeta^2$ is from the conversion 
between ${\bf k}_1$ and ${\bf P}_{H}$. The indices $i$ and $j$ denote, respectively, the 
flavors of the initial parton in the proton and of the outgoing fragmenting parton, and a sum over j and i are
implied. The 
incoming and outgoing parton momenta $p$ and $k_1$ satisfy $p=\xi P$ and 
$k_1=P_H/\zeta$. (Indices $i$ and $j$ for incoming and outgoing partons $p_i$ and 
$k_{1,j}$ will not be shown explicitly on the momenta but are understood).
$f_{i/P}(\xi;\mu)$ and $d_{H/j}(\zeta;\mu)$ are the collinear parton distribution and 
fragmentation functions respectively, with a renormalization group scale $\mu$. It
is also useful to define partonic variables 
\begin{equation}
\hat{x}\equiv \frac{Q^2}{2p\cdot q} = \frac{x}{\xi} \, , \qquad \hat{z} \equiv \frac{k_1\cdot p}
{p\cdot q} = \frac{z}{\zeta} \, , \qquad  \Tsc{k}{1}\equiv \frac{\Tsc{P}{H,}}{\zeta} \, .
\end{equation}
Note that at large transverse momentum, the cross section starts at
order $\order{\alpha_s}$ which is  finite and well-behaved for $\Tsc{q}{} > 0$.
The possible kinematical scenarios at $\Tsc{q}{} \sim Q$ for the 
partonic sub cross section in the integrand on the right-hand side of \eref{xsecfact} are 
sketched in \fref{basickinematics}. The dashed lines represent generic parton momenta in 
that all are assumed to be massless and highly non-collinear, but the exact identities of 
the partons are left unspecified. 

  The large $q_{\rm T}$ factorized cross section in
  \eref{xsecfact} has power corrections suppressed by $1/\Tscsq{q}{}$, or $1/Q^2$ 
  when $\Tsc{q}{} = \order{Q}$. 
  Those corrections are not negligible in the small $\Tsc{q}{}$ limit.
  In that limit, the cross section is best described in terms 
  of TMD factorization, wherein the power corrections are $O(\Tscsq{q}{}/Q^2)$. 
  The first term in \eref{xsecfact} contains contributions that would be counted as power suppressed 
  in the $\Tsc{q}{}/Q \to 0$ limit, so 
  \eref{xsecfact} cannot generally be inferred from the high transverse
  momentum behavior of TMD factorization results.

Since transverse momentum is frame dependent, characterizing its size requires some 
clarification.  In light of the outline of regions in \sref{intro}, we must ask what criteria generally need to be satisfied for a transverse 
momentum to be considered large or small. 
For this, define
\begin{equation}
 k \equiv k_1 -q  \, .
\end{equation}
The $k$ momentum would be the target parton momentum in the small transverse 
momentum limit where $k^2 \approx 0$ and the parton model $2 \to 1$ subprocess 
$\gamma^* q \to q$ applies. (See \fref{hand}.) Note that in \fref{basickinematics}(a), 
however, all the final state particles are at wide angles relative to one another, so internal 
propagators are off shell by order $Q^2$. It will also be useful to define
\begin{equation}
k_X \equiv p + q - k_1 \, .
\end{equation}
Note that in \fref{basickinematics}(b) $k_X^2 = 0$. 

Within the blob in \fref{basickinematics}(a), two basic forms of propagator denominators 
may arise:
\begin{align}
&{} \frac{1}{k^2 + \order{\Lambda_{\rm QCD}^2}} \, , \label{e.denomsizes1} \\ 
&{} \frac{1}{k^2 + \order{Q^2}} \, . \label{e.denomsizes2}
\end{align}
In \eref{denomsizes1}, the $\order{\Lambda_{\rm QCD}^2}$ terms are very small mass 
scales associated with non-perturbative physics. Equation~\eqref{e.denomsizes2} involves 
either the virtual photon vertex or emissions corresponding to (the wide-angle) $k_X$.
Note that 
$k \cdot q \sim q \cdot p = \order{Q^2}$. 
These propagator denominators illustrate 
the sort of power counting arguments necessary to justify the relevant factorization 
approximations in each region of partonic kinematics outlined in the introduction. 
For instance, when $|k^2| \approx \Lambda_{\rm QCD}^2$, the $k$-line is nearly on-shell 
and collinear to the target proton. So it should then be considered part of the proton blob, 
and the relevant physical picture becomes the handbag topology in \fref{hand} with $k$ 
now playing the role of the target parton. The $\Tscsq{q}{}/Q^2 \ll 1$ approximations that 
lead to TMD factorization apply here. Namely, the $k^2$ in \eref{denomsizes2} can be 
neglected relative to the $\order{Q^2}$ terms, although no small $k^2$ approximations are 
appropriate for \eref{denomsizes1}.

When $|k^2| \approx Q^2$, the $\order{\Lambda_{\rm QCD}^2}$ terms can be neglected in 
\eref{denomsizes1} and all of the blob in \fref{basickinematics}(a) can be calculated in 
pQCD with both $Q^2$ and $k^2$ acting as hard scales. Of course, it is then no longer 
appropriate to neglect $k^2$ relative to $Q^2$ terms in \eref{denomsizes2}, so this is 
the large transverse momentum region 3.

Explicit diagrammatic calculations, keeping small masses, easily confirm that the 
coefficients of the $\order{Q^2}$ and $\order{\Lambda_{\rm QCD}^2}$ terms in 
\erefs{denomsizes1}{denomsizes2} are simple numerical factors not radically different from 
1. Moreover, this generalizes to entire diagrams including propagator numerators. 
Thus, the ratio $|k^2|/Q^2$ is the relevant Lorentz invariant measure of the size of 
transverse momentum, in the sense that it should be much less than $1$ for the small 
transverse momentum approximations to be accurate. Calculating it in terms of $\hat{z}$, $
\Tscsq{q}{}$ and $Q^2$,
\begin{equation}
\left| \frac{k^2}{Q^2} \right| = 1-\hat{z} + \hat{z} \frac{\Tscsq{q}{}}{Q^2} \, . \label{e.rattm}
\end{equation}  
For current fragmentation, $z$ is fixed at some value not too much smaller than $1$. (In 
practice it is often assumed to be between approximately $0.2$ and $0.8$.)
In the integral over $\zeta$, $z < \hat{z} < 1$ so 
\begin{equation}
\frac{\Tscsq{q}{}}{Q^2} < \left| \frac{k^2}{Q^2} \right| <  1 - z \parz{1 - \frac{\Tscsq{q}{}}
{Q^2}} \, ,  \label{e.rat2}
\end{equation}
assuming $\Tsc{q}{} < Q$.
So, for any $z$ in the current region, $\Tscsq{q}{}/Q^2 \ll 1$ signals the onset of the TMD 
factorization region while $\Tscsq{q}{}/Q^2 \sim 1$ signals the onset of the large transverse 
momentum region where fixed order pQCD is optimal. That is, $\Tscsq{q}{}/Q^2 \ll 1$ 
implies region 1 or 2 of the introduction, while $\Tscsq{q}{}/Q^2 \gtrsim 1$ implies region 3. This 
establishes that it is the magnitude of $\T{q}{}$ specifically, defined in \eref{qtdef}, that is 
most useful for assessing the transition between different regions.

Another way to estimate the boundary between large and small transverse momentum is 
to recall that the region 1 and 2 methods of calculating are the result of a small $\Tsc{q}{}$ 
approximation. Thus, one may examine the effect of that approximation 
in specific fixed order calculations. For SIDIS at $\order{\alpha_s}$ in the small 
$\Tscsq{q}{}/Q^2  \to 0$ limit, the cross section is proportional to
\begin{equation}
d(z;\mu) \int_x^1 \frac{\diff{\xi}}{\xi} f(\xi;\mu) P(x/\xi) 
+ f(x;\mu) \int_z^1 \frac{\diff{\zeta}}{\zeta} d(\zeta;\mu) P(z/\zeta) 
+ 2 C_F f(x;\mu) d(z;\mu) \parz{\ln \parz{\frac{Q^2}{\Tscsq{q}{}}} - \frac{3}{2}} \, . 
\label{e.asym}
\end{equation}
where the functions $P$ are lowest order splitting functions and sums over flavors are implied. 
(See, for example, Eq.~(36) of~\cite{Nadolsky:1999kb}.) The appearance of the logarithm 
is a consequence of approximations specific to the small $\Tsc{q}{}$ limit. The actual fixed 
order calculation of the cross section from which this is obtained is positive everywhere. Therefore, the 
$\Tscsq{q}{}/Q^2  \to 0$ approximation is surely 
inappropriate once $\Tscsq{q}{}$ is so large that the cross section as calculated in 
\eref{asym} becomes significantly negative. If the first two terms of \eref{asym} are of order unity or 
less, then this happens when $\Tsc{q}{} \gtrsim Q$.  One can use \eref{asym} to estimate 
where region 3 methods are definitely needed. Specific calculations in 
\cite{Boglione:2014oea} show values of $\Tsc{q}{}$ above which the cross section goes 
negative in typical calculations. For example, Fig.~1 of \cite{Boglione:2014oea} suggests 
that the change of sign occurs before $\Tsc{q}{} \sim Q/2$ in typical SIDIS kinematics. 
(This  further establishes $\Tscsq{q}{}/Q^2$ as the relevant transverse momentum ratio.)

Another question is whether, for a particular combination of $x$ and $z$, the large 
$\Tsc{q}{}$ calculation should be expected to be well-described by the $\order{\alpha_s}$ 
calculation or whether higher orders are needed. For the leading $\order{\alpha_s}$ large 
transverse momentum cross section, the partonic process is $2 \to 2$ with all partons 
massless and on-shell (\fref{basickinematics}(b)), i.e., $k_X^{2\,\,(\alpha_s)} = 0$. For the 
$\order{\alpha_s}$ calculation to be a good approximation, therefore, the ratio 
$k_X^2/Q^2$ (\fref{basickinematics}(a)) must be small enough that it does not affect the $k^2$ terms in 
\eref{denomsizes1} and \eref{denomsizes2}. Considering $k^2/Q^2$ but now in terms of $k_X^2$ 
instead of $\hat{z}$,
\begin{align}
	\left|\frac{k^2}{Q^2} \right| = 
  \frac{1}{1-\hat{x}+\hat{x}q^2_{\rm T}/Q^2}
  \left[
    \frac{q_{\rm T}^2}{Q^2} 
    + \hat{x}\frac{k_X^2}{Q^2}\left(1-\frac{q^2_{\rm T}}{Q^2}\right)
  \right] \, .
\label{e.ksq}
\end{align}
So the $k^2$ terms in \eref{denomsizes1} and \eref{denomsizes2} are nearly independent of 
$k_X$ if $k_X^2 /Q^2 \ll 1$.
Otherwise, higher orders in 
$\alpha_s$ are necessary to generate the non-zero $k_X^2$. In terms of $\hat{x}$, $\hat{z}$ and 
$Q^2$,
\begin{equation}
\frac{k_X^2}{Q^2} = \frac{(1-\hat{x})(1-\hat{z})}{\hat{x}} - \hat{z} \frac{\Tscsq{q}{}}{Q^2} \, . 
\label{e.kXsq}
\end{equation}
In practice, typical $\hat{x}$ and $\hat{z}$ are largely determined by the distributions in 
longitudinal momentum fraction in PDFs and FFs for a particular kinematical scenario.  If it 
turns out that they are mostly dominated by moderate values of $\hat{x}$ and $\hat{z}$, 
then the $k_X^2/Q^2 \ll 1$ criterion is not difficult to satisfy for large $\Tsc{q}{} \sim Q$. 
Then, the leading order in $\alpha_s$ can reasonably be expected to dominate at large 
transverse momentum. If, however, the typical $\hat{x}$ and $\hat{z}$ are much smaller 
than 1, then they force a large average $k_X^2$. This can create the situation that  
higher order corrections are larger relative to $\order{\alpha_s}$ for certain regions of 
transverse momentum. (If there is large sensitivity to the kinematical threshold at $k_X^2 \approx 0$, then 
this can also induce large higher order corrections.)

Note that the shape of the transverse momentum dependence can be significantly affected 
by the PDFs and FFs because of the correlation between $\zeta$ and $\xi$:
\begin{equation}
\zeta = z \parz{ \frac{\xi - x + x \Tscsq{q}{}/Q^2}{\xi - x  - x k_X^2/Q^2} } \, .
\end{equation}
Inclusive quantities are sensitive to the peak in the cross section at small $\Tsc{q}{}$ (and 
small $k_X^2$) and thus are mainly sensitive to the region where $\zeta \approx z$. Both 
large $\Tscsq{q}{}$ and large $k_X^2$, however, push $\zeta$ to values significantly 
higher than $z$. 

\section{Existing Measurements and Calculations}
\label{s.results}

Given the discussion above, we should expect to find reasonable agreement between fixed order 
calculations and SIDIS measurements where $\Tsc{q}{}/Q$ ratios easily exceed 1 and typical $x$ and 
$z$ are not such that higher orders are extremely large. This at first appears promising when 
considering H1~\cite{Aktas:2004rb} kinematics~\footnote{H1 is the experimental collaboration that produced the data in~\cite{Aktas:2004rb}.}, where the fixed order large-$\Tsc{q}{}$
prediction from Fig. 4 of~\cite{Daleo:2004pn} (copied here in
\fref{dal}) gives a satisfactory description of $\pi^0$
production data if $\order{\alpha_s^2}$ corrections are
included. An obvious concern is that the order-of-magnitude higher order
corrections needed might be signaling a breakdown of perturbative
convergence.  But as explained in~\cite{Daleo:2004pn}, this behavior
is most likely due simply to the particular kinematics of the H1 data. 
Indeed $5\times10^{-5} \lesssim x \lesssim 5\times10^{-3}$ for the
data in~\fref{dal}. Also, the cross section is integrated over $z$ with 
\begin{align}
	z=\frac{P \cdot P_{\pi}}{P \cdot q}= \frac{2x E_p^2}{Q^2} (E_{\pi^0}/E_p) (1-\cos(\theta)) 
\end{align}
where $E_p$, $E_{\pi^0}$ are the energies of the proton and final 
state $\pi^0$ and $\theta$  is the polar scattering angle
relative to the incoming proton direction, all defined in the H1
laboratory frame. The data are constrained to $E_{\pi^0} / E_p > 0.01$ and 
$5^{\circ}< \theta < 25^{\circ}$. Using H1 kinematics, we find that the $z$ values included in~\fref{dal} can be as small as $\sim 0.001$. 
See section IV of~\cite{Daleo:2004pn} for more on the role of H1 cuts in producing \fref{dal}. A calculation similar to~\cite{Daleo:2004pn} was performed in~\cite{Kniehl:2004hf} with similar results.
The small values of $z$ are cause for caution. This is a region where a description of non-perturbative properties in terms of fracture functions 
(see, e.g.,~\cite{Trentadue:1993ka,Kotzinian:2013kra} and also applications to SIDIS in~\cite{Daleo:2003xg,Daleo:2003jf}) is more appropriate.

One consequence of the small $z$ values is that $\Tsc{q}{} = \Tsc{P}{H,}/z$ is very
large for each $\Tsc{P}{H,}$ point in \fref{dal}, so even the smallest
transverse momenta in the plots correspond to very large transverse
momenta by the criterion of \eref{rat2}. So it is maybe reasonable to expect that
the full range of transverse momentum observable in \fref{dal}
corresponds to region 3 large transverse momentum.  Also, for most
of the range of the integrals over $\zeta$ and $\xi$ in
\eref{xsecfact}, $\hat{x}$ and $\hat{z}$ are close to zero, but with a large 
contribution at $\hat{x} \approx 1$.  Given
\erefs{ksq}{kXsq}, therefore,  it is perhaps not surprising that
$\order{\alpha_s^2}$ calculations actually dominate since they are
needed to produce the large $k_X^2/Q^2$. 

The question then, however, is whether fixed order SIDIS calculations continue to be in 
reasonable agreement with measurements at more moderate $x$ and $z$ and at large 
$\Tsc{q}{}$, where the expectation is that agreement should improve, at least
with the inclusion of $\order{\alpha_s^2}$ corrections.
Figure~\ref{f.dalcomp}, shows that this is not the case, however. 
The order 
$\order{\alpha_s}$ and $\order{\alpha_s^2}$ curves are obtained with 
 an analogous computer calculation as that used in 
Ref.~\cite{Daleo:2004pn} to generate \fref{dal}, but modified to be consistent with the 
kinematics of the corresponding experimental data. (We have verified that the curves in 
\fref{dal} are reproduced.) The data are from recent COMPASS measurements for 
charged hadron production~\cite{Aghasyan:2017ctw}. Neither leading order nor next-to-leading order calculations 
give reasonable agreement with the measurements, even for moderate $x$, $z$ and 
$\Tsc{q}{} > Q$, as both systematically undershoot the data, most significantly at the more 
moderate values of $x$ close to the valence region. 
\begin{figure}
\includegraphics[scale=0.85]{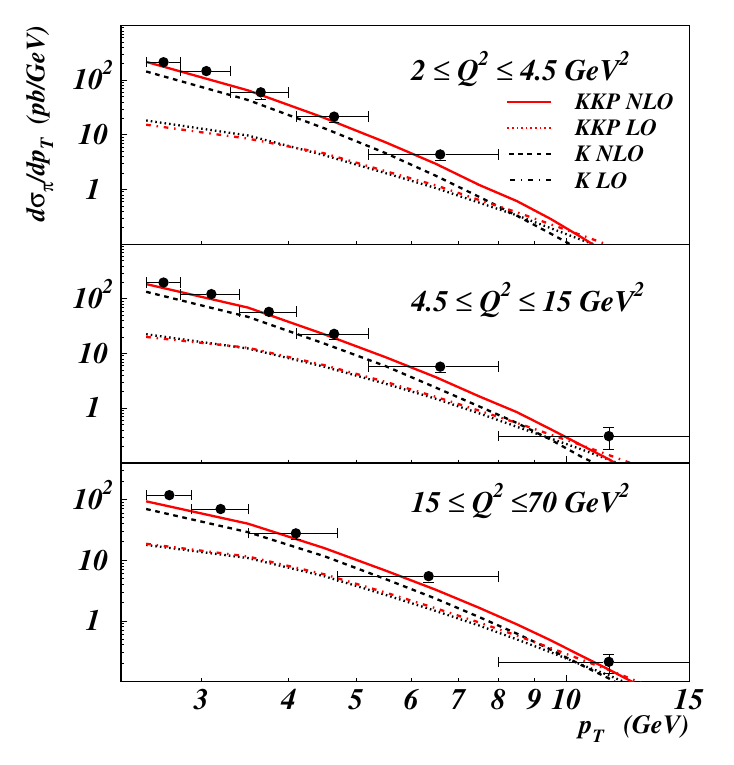}
\caption{Fig.~4 from~\cite{Daleo:2004pn}. The differential cross section was integrated 
over $x$, $z$ and bins of $Q$ with H1 cuts,
 calculated with both leading order and next-to-leading order, and compared with 
$\pi^0$ production data from~\cite{Aktas:2004rb}. Here $p_T$ corresponds to our 
$\Tsc{P}{H,}$ -- see \eref{qtdef}. Note the large correction from $\order{\alpha_s^2}$.}
\label{f.dal}
\end{figure}
\begin{figure}
\includegraphics[width=0.75\paperwidth]{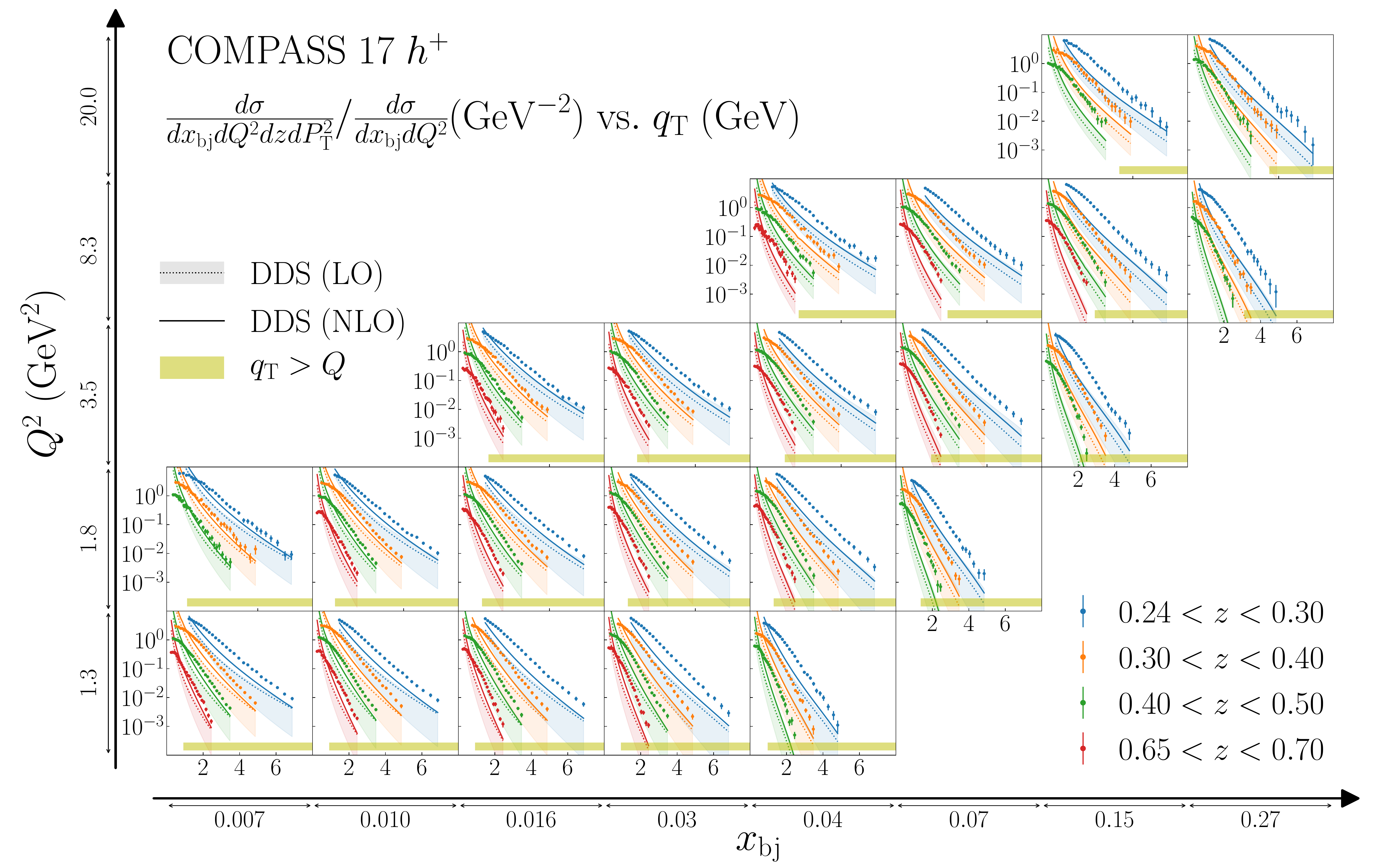}
\caption{Calculation of $O(\alpha_s)$ and $O(\alpha_s^2)$ transversely differential 
multiplicity using code from~\cite{Daleo:2004pn}, shown as the curves labeled DDS (for Daleo-de Florian-Sassot). The 
bar at the bottom marks the region where $\Tsc{q}{} > Q$. The PDF set used is 
CJNLO~\cite{Accardi:2016qay} and the FFs are from~\cite{deFlorian:2007ekg}. 
Scale dependence is estimated using 
$\mu=\left((\zeta_{Q} Q)^2+(\zeta_{q_{\rm T}}q_{\rm T})^2\right)^{1/2}$ where
the band is constructed point-by-point in $q_{\rm T}$ by taking the min and max of the 
cross section evaluated across the grid 
  $\zeta_{Q}\times \zeta_{q_{\rm T}}=[1/2,1,3/2,2]\times[0,1/2,1,3/2,2]$
  except $\zeta_{Q}=\zeta_{q_{\rm T}}=0$. The red band is generated
  with $\zeta_{Q}=1$ and $\zeta_{q_{\rm T}}=0$.  A lower bound of $1$~GeV is 
place on $\mu$ when $Q/2$ would be less than $1$~GeV. }
\label{f.dalcomp}
\end{figure}
At smaller $x$ the disagreement lessens, as might be expected given the trend in 
\fref{dal}. To highlight the valence region ($x \gtrsim 0.1$) at the larger values of $Q$, we 
have plotted the ratio between data and theory in \fref{compassrat} for three particular 
kinematic bins from \fref{dalcomp}. Even including the $\order{\alpha_s^2}$ 
correction, the 
deviation is typically well above a factor of $2$, even for $\Tsc{q}{}$ significantly larger than 
$Q$.
In this context it is also worth considering Fig.~8 of \cite{Kniehl:2004hf}, which is for similar kinematics to \fref{dal} but for charged 
hadrons measured at ZEUS~\cite{Derrick:1995xg}. The next-to-leading order $K$ factor is $\gtrsim 1.5$ for large transverse momentum.
\begin{figure}
\includegraphics[width=0.6\paperwidth]{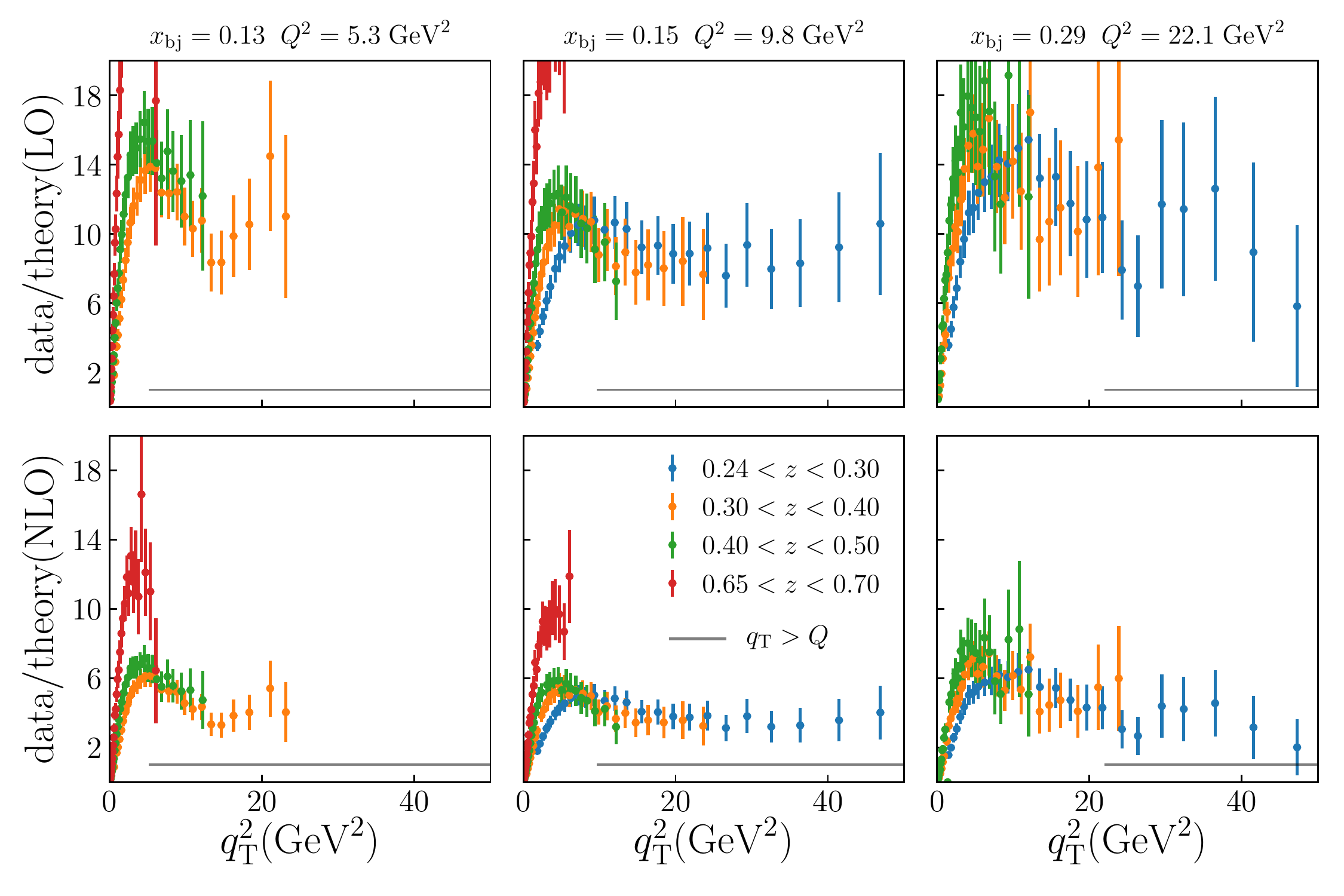}
\caption{Ratio of data to theory for several near-valence region panels in \fref{dalcomp}. 
The grey bar at the bottom is at $1$ on the vertical axis and marks the region where $\Tsc{q}{} > Q$.}
\label{f.compassrat}
\end{figure}
\begin{figure}
\includegraphics[width=0.7\paperwidth]{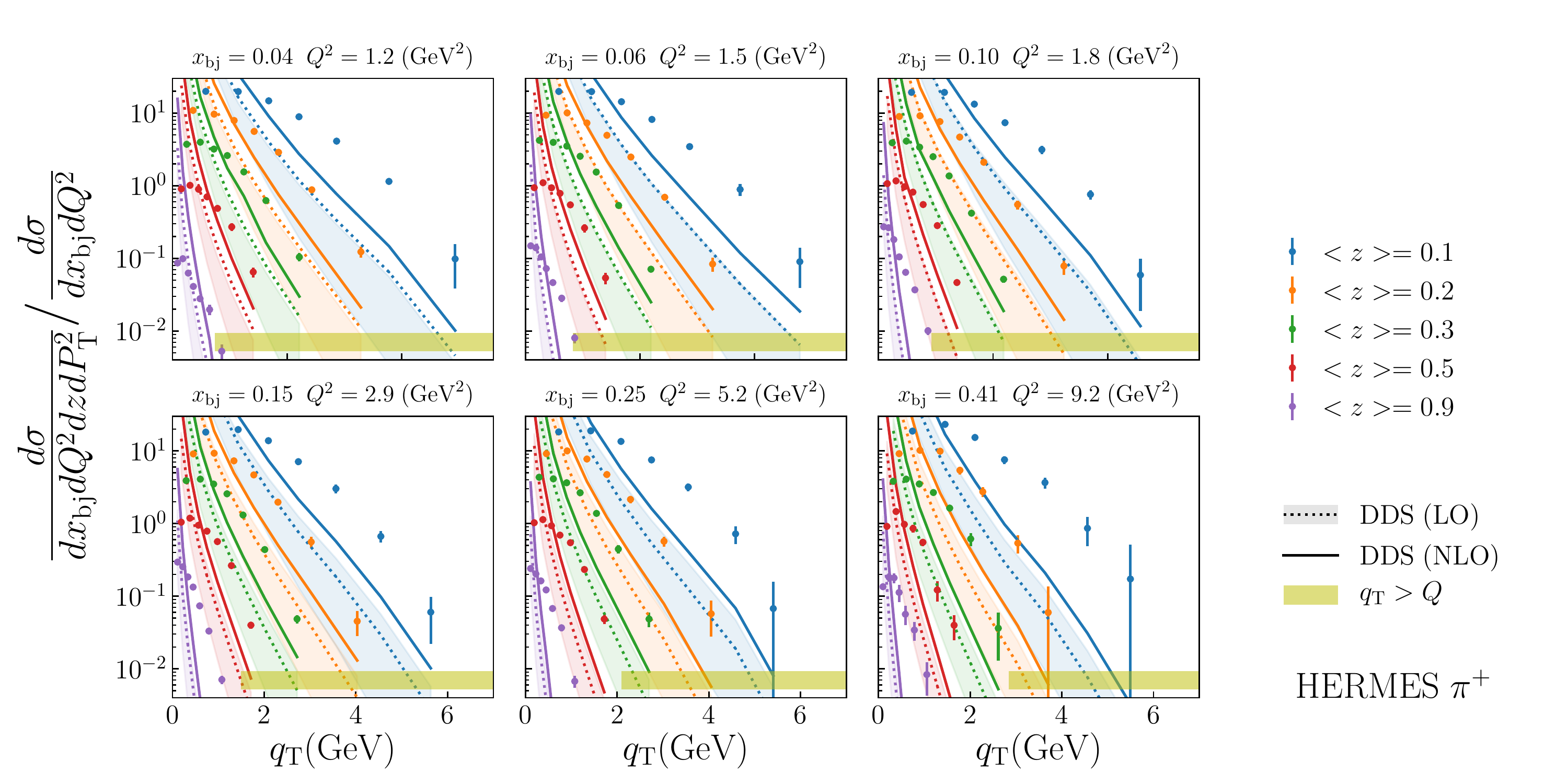}
\caption{Calculation analogous to \fref{dalcomp} but for $\pi^+$ production measurements 
from~\cite{Ma:2015lka}. }
\label{f.dalcompherm}
\end{figure}
\begin{figure}
\includegraphics[width=0.55\paperwidth]{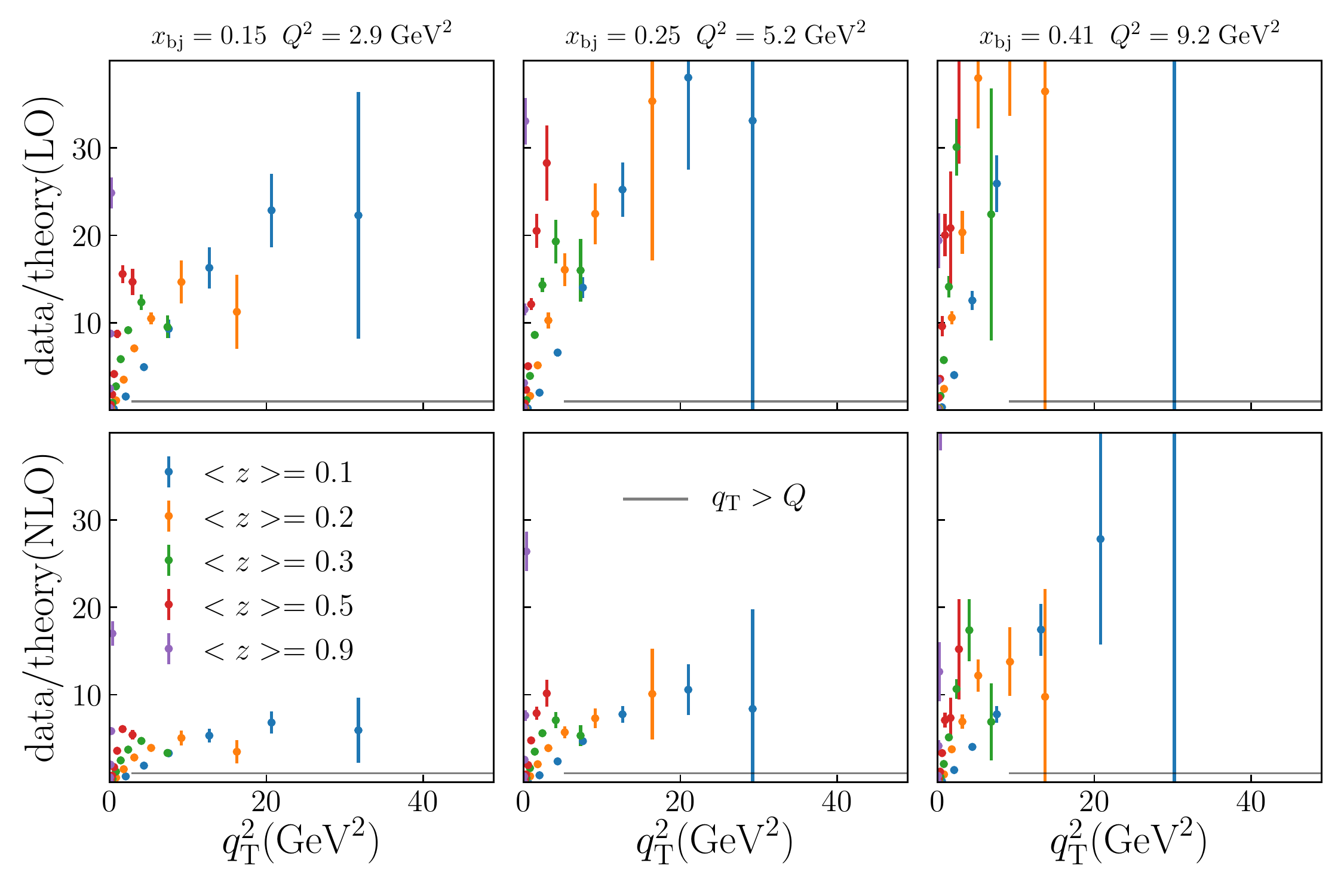}
\caption{Ratio calculation analogous to \fref{compassrat} but for $\pi^+$ production 
measurements from~\cite{Ma:2015lka}.}
\label{f.dalcomphermrat}
\end{figure}
At least one other set of SIDIS data at somewhat different kinematics exhibits the same 
trend. This is the set of HERMES measurements of $\pi^+$ multiplicities~\cite{Ma:2015lka} 
shown in \fref{dalcompherm}. Note that the kinematics very much correspond to the 
valence region for the target. Figure~\ref{f.dalcomphermrat} shows that the failure to match 
the data is even more pronounced than in the COMPASS case. Even for $Q > 3$~GeV and 
$\Tsc{q}{} > Q$, the difference is nearly an order of magnitude.

\section{Discussion}
\label{s.conc}

We have argued that there is tension between existing fixed order pQCD calculations and 
at least two sets of large transverse momentum measurements where those calculations 
should be reasonably accurate, and that this disagreement is too large to be attributable to 
$\Tsc{q}{}$ being too small. 
  Thus, it appears to us to be a genuine mystery that needs attention,
  especially for TMD phenomenology.
The TMD formalism relies on approximations that apply only in the $\Tsc{q}{}/Q \to 0$ limit, so it 
is critical to have an alternative approach to describe the transition to very large transverse 
momentum. If standard fixed order collinear pQCD is not adequate for this, then something 
new is needed.
 
It is worth pointing out that one encounters similar problems in Drell-Yan scattering, where 
a lowest order calculation with current PDF sets is easily found to undershoot the lowest 
available $Q$ data by very large factors. It is less clear how to interpret the disagreement 
here, however, since most of the existing data for lower $Q$ regions are close to the 
threshold region and including threshold resummation introduces extra subtleties. 

The observations of this article have focused on unpolarized cross sections, but the implications 
extend to spin and azimuthally dependent cross sections, since the key issue is the relevance of different 
types of transverse momentum dependence.
 
There are a number of possible resolutions that deserve further investigation.  
An interesting one is that the hadronization mechanism is different in high-transverse-
momentum SIDIS from the usual picture in terms of universal FFs. Models used in Monte 
Carlo event generators might be a source of ideas regarding this possibility. In the context of this possibility, 
it is noteworthy that much of the data for SIDIS transverse momentum dependence is describable in a Gaussian 
model of TMDs~\cite{Anselmino:2013lza,Signori:2013mda}.
In pQCD, there are also arguments that certain higher twist correlation functions actually 
dominate over leading twist functions. In this picture, the $q \bar{q}$ pair that 
ultimately forms  the final state is directly involved in the hard 
part~\cite{Ma:2015lka,Kang:2014pya}. 

It is possible that threshold effects are important
~\cite{deFlorian:2013taa,Westmark:2017uig}. If that is the case, then there are serious implications for 
TMD studies, because additional non-perturbative effects beyond those 
associated with intrinsic transverse momentum can then be important ~\cite{Collins:2007ph}. However, the largest 
$x$ and $z$ in \fref{compassrat} and \fref{dalcomphermrat}, where the disagreement is the 
worst, corresponds to valence regions of $x$ and $z$, well away from kinematics where 
partonic kinematics are close to kinematical thresholds for most $\xi$ and $\zeta$. 
Furthermore, this would contradict the observation in \fref{dal} that fixed order corrections 
alone succeed in describing data, even for very small $x$ and $z$.

Another possibility is that FFs and/or PDFs are not well-enough constrained to handle 
the particular kinematical scenarios that arise at large $\Tsc{q}{}$. 
To see that this is plausible, consider the large $\Tsc{q}{}$ observable
\begin{equation}
\frac{\diff{\sigma}{} }{\diff{Q^2} \diff{x} \diff{z} \diff{\Tscsq{q}{}}} \sim \int_{\xi_{\rm min}}^1 
\diff{\xi}{} f\parz{\xi} d\parz{\zeta=z \parz{1 + \frac{x \Tscsq{q}{}}{(\xi - x) Q^2}}} \, , \label{e.ob1}
\end{equation}
where the ``$\sim$'' is to indicate the combination of PDF and FF that appear in fixed order 
pQCD calculations at large transverse momentum and lowest order. The minimum final 
state momentum fraction is 
\begin{equation}
\xi_{\rm min} = x \parz{1 + \frac{z \Tscsq{q}{} }{(1 - z)Q^2}} \, .
\end{equation}
Note that small $\xi$ ($\xi \approx x$) tends to select the large $\zeta \gg z$ region and 
vice versa, and the nature of the correlation between $\zeta$ and $\xi$ changes as 
$\Tsc{q}{}$ varies. Also, very large $\Tsc{q}{}$ forces $\xi_{\rm min}$ to be large. 
Therefore, while the hard scattering gives a characteristic power-law shape to the large 
$\Tsc{q}{}$ dependence, both its shape and normalization are also significantly influenced 
by the large $\xi$ and $\zeta$ behavior in the collinear PDFs and FFs. Note also that gluon 
PDFs and FFs appear in the first non-vanishing order in \eref{ob1}.

Contrast that with the observable,
\begin{equation}
\sum_{\text{Hadron Flavors}} \int z \diff{z} \diff{\Tscsq{q}{}} \frac{\diff{\sigma}{} }{\diff{Q^2} 
\diff{x} \diff{z} \diff{\Tscsq{q}{}}} \
= \frac{\diff{\sigma}{} }{\diff{Q^2} \diff{x}} \sim f\parz{x} \, , \label{e.ob2}
\end{equation}
that is, the total DIS cross section; or with the $\Tsc{q}{}$-integrated SIDIS cross section,
\begin{equation}
\int \diff{\Tscsq{q}{}} \frac{\diff{\sigma}{} }{\diff{Q^2} \diff{x} \diff{z} \diff{\Tscsq{q}{}}} \
= \frac{\diff{\sigma}{} }{\diff{Q^2} \diff{x} \diff{z}}  \sim f\parz{x} d\parz{z} \, . \label{e.ob3}
\end{equation}
In \eref{ob2}, the dominant contribution at leading order is from quark PDFs only, and is 
evaluated only at a single value $x$. Likewise, \eref{ob3} is only sensitive to the FF at 
$\zeta = z$.  Of course, \eref{ob1}, \eref{ob2}, and \eref{ob3} are all for basically the 
same process, but in going from \eref{ob1} to \eref{ob2} or \eref{ob3} information is lost 
in the integrations and summations, and the sensitivity to the PDFs and FFs is 
consequently less detailed. Most typically, however, low-$Q$ fits that are aimed at constraining the 
valence region at lower $Q$ use observables like \eref{ob2} or \eref{ob3}. But \eref{ob1} scans through 
small to large values of $\xi$ and $\zeta$ as $\Tsc{q}{}$ varies. The question then arises 
whether existing fits maintain enough information to predict \eref{ob1} observables reliably. 
Note that it has already been suggested in the past~\cite{Berger:1998ev} to use 
transversely differential Drell-Yan measurements at large $\Tsc{q}{}$ and including 
smaller $Q$ to constrain gluon distributions. An analogous possibility applies to 
gluon FFs at large $\zeta$ in large transverse momentum SIDIS or back-to-back 
hadron pair production in $e^+ e^-$ annihilation. To test this, it would be informative to 
include the large transverse momentum behavior of lower-$Q$ but highly differential cross sections 
in global simultaneous fits of PDFs and FFs. It is also noteworthy that in very early 
calculations~\cite{Halzen:1978et} that gave rise to satisfactory fits in Drell-Yan scattering, 
the gluon and sea distributions used were much larger than modern ones. This suggests 
that contributions from gluon and sea PDFs and FFs at larger values of $x$ may need to 
be reassessed in light of the mismatches above. 

We leave further investigation of all these possibilities to future work.  A resolution is an 
important part of the goal to understand SIDIS generally in terms of an underlying partonic 
picture.

 
\begin{acknowledgments}	
We thank J. Collins, L. Gamberg, F. Halzen, J. Owens, J.-W. Qiu, and W. Vogelsang for very useful 
discussions.  We thank R. Sassot for explanations regarding the code in \cite{Daleo:2004pn}. 
T.~Rogers's work was supported by the U.S. Department of Energy, Office of 
Science, Office of Nuclear Physics, under Award Number DE-SC0018106. This work was 
also supported by the DOE Contract No. DE- AC05-06OR23177, under which Jefferson 
Science Associates, LLC operates Jefferson Lab. N.~Sato was supported by DE-FG-04ER41309.
B.~Wang was supported in part by the National Science Foundation of China 
(11135006, 11275168, 11422544, 11375151, 11535002) and the Zhejiang University Fundamental
Research Funds for the Central Universities (2017QNA3007). J.~O.~Gonzalez-Hernandez work was partially supported
by Jefferson Science Associates, LLC under  U.S. DOE Contract \#DE-AC05-06OR23177 and by the 
U.S. DOE Grant \#DE-FG02-97ER41028.
\end{acknowledgments}

\bibliography{bibliography}

\end{document}